\documentclass[preprint,
 aip,
jcp,
 jmp,%
 amsmath,amssymb,
]{revtex4-1}

\usepackage{graphicx}
\usepackage{dcolumn}
\usepackage{bm}
\usepackage{enumerate}
\usepackage{float}
\usepackage{subfig}
\usepackage{wrapfig}
\usepackage[justification=centerlast]{caption}
\begin{document}


\title[]{Micro-Optomechanical Movements (MOMs) with SoftOxometalates (SOMs): Controlled Motion of Single Soft-Oxometalate Pea-pods Using Exotic Optical Potentials}


\author{Basudev Roy}
 \affiliation{Department of Physical Sciences, IISER-Kolkata, Mohanpur 741252, India}
\author{Atharva Sahasrabudhe}%
\affiliation{EFAML, Materials Science Centre, Department of Chemical Sciences, IISER-Kolkata, Mohanpur 741252, India}%
\author{Bibudha Parasar}
\affiliation{EFAML, Materials Science Centre, Department of Chemical Sciences, IISER-Kolkata, Mohanpur 741252, India}
\author{Nirmalya Ghosh} \email{nghosh@iiserkol.ac.in}
\affiliation{Department of Physical Sciences, IISER-Kolkata, Mohanpur 741252, India}
\author{Prasanta Panigrahi}
\affiliation{Department of Physical Sciences, IISER-Kolkata, Mohanpur 741252, India}
\author{Ayan Banerjee} \email{ayan@iiserkol.ac.in}
\affiliation{Department of Physical Sciences, IISER-Kolkata, Mohanpur 741252, India}
\author{Soumyajit Roy}
 \email{s.roy@iiserkol.ac.in}
\affiliation{EFAML, Materials Science Centre, Department of Chemical Sciences, IISER-Kolkata, Mohanpur 741252, India}


\begin{abstract}
An important challenge in the field of materials design and synthesis is to deliberately design mesoscopic objects starting from well-defined precursors and inducing directed movements in them to emulate biological processes. Recently, mesoscopic metal-oxide based Soft Oxo Metalates (SOMs) have been synthesized from well-defined molecular precursors transcending the regime of translational periodicity. Here we show that it is actually possible to controllably move such an asymmetric SOM-with the shape of a `pea-pod' along complex paths using tailor-made sophisticated optical potentials created by spin-orbit interaction of light due to a tightly focused linearly polarized Gaussian beam propagating through a stratified medium in an optical trap. We demonstrate motion of individual trapped SOMs along circular paths of more than 15 $\mu$m in a perfectly controlled manner by simply varying the input polarization of the trapping laser. Such controlled motion can have a wide range of application starting from catalysis to the construction of dynamic mesoscopic architectures. 

\end{abstract}

\maketitle
\section{\label{intro}INTRODUCTION}
\noindent 
A long cherished goal of scientists has been to emulate transport phenomena involved in life-processes under conditions far from equilibrium. Living systems accomplish this feat by using motor proteins to actively transport ingredients over large distances \cite{Bruce}. Synthetically emulating such a process would involve two steps: (1) Controlled generation of mesoscopic objects starting from well-defined precursors; (2) Using physical means to induce controlled motion in such mesoscopic objects.  In recent times, a class of stimuli-responsive metal-oxide based mesoscopic materials, called soft-oxometalates or SOMs \cite{sou11} have been synthesized - these respond to  variations in external fields, and are endowed with soft-matter properties. It is hence reasonable to envisage that if a SOM that has a responsive component to any external physical perturbation can be deliberately designed, a synthetic model system showing controlled motion comparable to biological systems can be constructed. More specifically, the motion of motor proteins such as kinesin and myosin which are involved in muscular movements can be emulated at a more simplistic level by controllably applying directed motion to a SOM particle. Moreover, since any chemical stimulus leads to mechanical motion in proteins -  one could, in principle, extract information about chemical processes in proteins by inducing controlled motion in them. In this context, the interaction of light and matter naturally presents itself as a first choice. Furthermore, the level of control can be extended if one could actually move the SOM in a complex pre-designed path by known amounts. Such a path was designed and created using optical forces, and an optically responsive SOM was made to move along that path to design our model system. The details of this design is presented in this paper. 

Optical tweezers, by virtue of their ability to confine single mesoscopic particles and apply controlled forces ranging from sub-pN to several hundred pN \cite{per08}, are an ideal candidate to induce controlled motion in SOMs. While translation of trapped SOMs linearly by translating the optical trap can be achieved by simple means such as moving the trapping beam or translation stage \cite{vis93}, or by even more innovative approaches such as using an asymmetric trapping beam \cite{moha05}, translation along more complex paths which may be required to emulate biological processes more accurately, is not easily achievable. A step towards this direction has been taken in recent times with the advent of holographic optical tweezers which have led to the transport of trapped dielectric objects along pre-designed paths \cite{gri06}. Thus, while holographic tweezers are rather ubiquitous currently, it is important to note that they are typically created by coupling a fixed computer generated interference pattern into the trapping region, and changing the particle trajectory or stopping the particle during its motion is somewhat non-trivial. In our scheme, the trapped particle is moved by changing the angle of polarization of the input linearly polarized trapping beam, which gives us complete control of the motion in the radial direction (transverse to the propagation direction of the laser) both in terms of stopping the particle or changing its velocity. We use the fact that asymmetric birefringent particles have a preferred symmetry axis that can line up with the direction of the intensity gradient produced by the trapping beam, and also design a rather exotic optical potential in our optical trap in order to induce controlled MOMs (Micro-optomechanical movements) on individual pea-pod shaped SOMs, which are specifically designed for our scheme. The pea-pods have ${\rm MoO_3}$ skin and contain ${\rm P_2Mo}$ spheres within -  which makes them comparable from a constitutional and length scale point of view to a protein. The exotic optical potential is achieved by an enhancement of the spin-orbit interaction of light affecting the electric field distribution inside the sample chamber \cite{roy12}. The enhanced spin-orbit interaction is achieved by using cover-slips (refractive index (RI) 1.575 and thickness 250 $\mu$m) having different RI and thicker than the conventional cover slips used in optical tweezers (RI 1.516, thickness 130 - 160 $\mu$m). 

\begin{figure*}[ht]
 \centering{\includegraphics[scale=0.6]{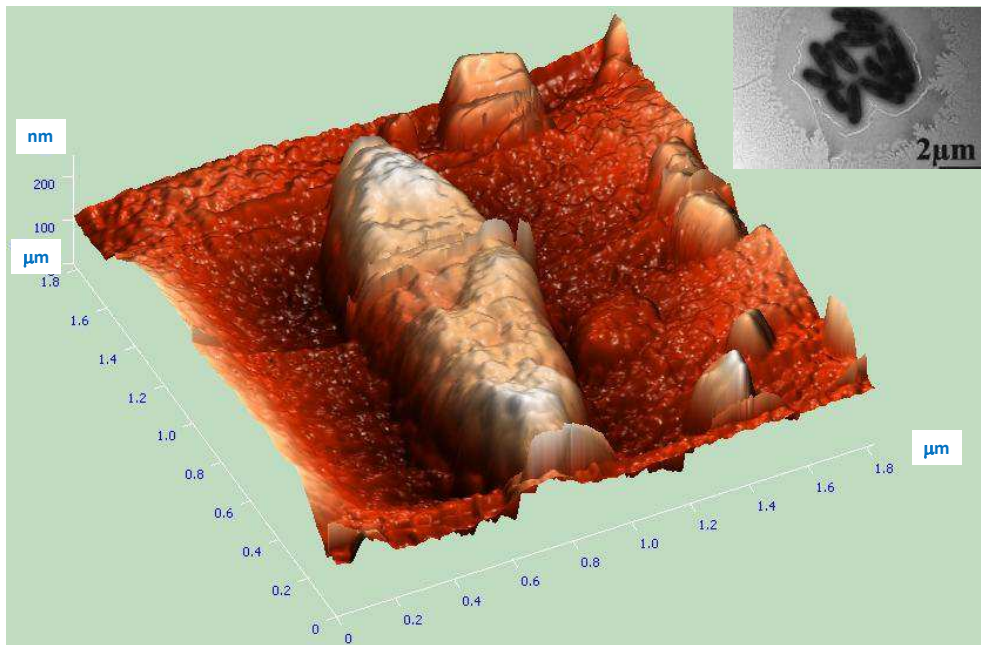}}
    \caption[Figure-1]{AFM image of a single pea-pod. Dimensions are in $\mu$m in the $x$ and $y$ axes, and in nm in the $z$-axis. Inset shows a bright field TEM image of a collection of pea-pods.}
\label{Figure-1}
\end{figure*}

While there exist several kinds of oxo-metalates \cite{sou11}, certain key features led us to choose the particular SOM used here. They are: 
\begin{enumerate}[(a).]
\item An inherent shape asymmetry - the SOM has a longitudinal dimension of 1-2 $\mu$m and a lateral dimension of around 500 nm (similar to a shape of a `pea-pod'). This would enhance form birefringence \cite{Born} which could, in principle, align the pea-pod with the electric field of the light in the trap. \item The design of the `pea-pod' is deliberated and known to  minute atomic detail and synthesized using a well-defined molecular precursor  \cite{sou11}. \item The `skin' of the `pea-pod' is composed of ${\rm MoO_3}$, a versatile immobilizing agent for diverse catalytic reactions as a catalyst carrier. \item The motion assigned to these SOM-pea-pods, could at least in principle, provide a means of deliberated information transfer from a single molecular level to micro- and macroscopic level using simple and controllable physico-chemical pathways comparable to biological systems. Such systems could be envisaged to be of use in controlled micro-reactors, drug-delivery and related nano-systems. These pea-pods have been characterized by an array of available techniques and their mesoscopic topology is captured in an AFM image in Fig. ~\ref{Figure-1}.
\end{enumerate}

The design of the optical potential on the other hand, is based on the well-known fact that tight focusing of a linearly polarized beam leads to a  space-dependent linear diattenuation at the focal region \cite{roh05,zeev07}, which gets amplified due to propagation of the tightly focused beam through a stratified medium. This leads to the intensity maxima near the focus being shifted from the beam axis to a ring of varying radius around the axis. Additionally, two lobes of higher intensity are formed inside the ring, where single particles could be preferentially trapped \cite{roy12}. The location of these local intensity maxima lobes can be continuously changed in the ring by changing the polarization of the trapping beam. We use this effect to trap pea-pods in either of the intensity maxima, and then drag the pea-pod along the ring periphery by changing the input polarization angle using a linear retarder ($\lambda/2$ waveplate). The motion induced in the pea-pod thus does not require any realignment of the trapping beam and therefore does not modify the strength of the potential encountered by the trapped particle. 

It is important to note that ring-like intensity structures in the focal plane are also observed using higher order optical beams such as Laguerre-Gaussian beams \cite{pad11}, and the optical potentials thus created can also spontaneously drag the particles along the ring periphery. However, similar to holographic tweezers, it is not possible to control the position of the particles in the ring using such beams since there exists a phase singularity in the beam center owing to which the phase vortexes around it -  a particle trapped in the ring would thus move spontaneously along the vortex. Stopping the particle would require switching the beam off, in which case the particle would no longer be trapped. This explains  our choice and design of the optical potential and that of SOM pea-pods. 
\section{\label{expmt}Materials and methods}
\subsection{\label{expapp}Description of optical tweezer system}
The setup is a standard optical tweezers configuration consisting of an inverted microscope (Carl Zeiss Axiovert.A1), and a trapping laser (Lasever LSR1064ML, wavelength 1064 nm, single transverse mode with maximum power 800 mW) coupled into the microscope back port, so that the laser beam is tightly focused on the sample using a high numerical aperture objective (Zeiss 100X, oil immersion, plan apochromat, 1.41 NA, infinity corrected) lens. About 25 $\mu$l of the sample consisting of pea-pods in water solution (1:10000 dilution) is placed in a sample chamber consisting of a microscope glass slide (1 mm thickness) and cover slip. The cover slip is made of a transparent polymer (Sigma Aldrich Hybri Cover-slips, Part no. Z365912-100EA) having refractive index (RI) of 1.575 (at 1064 nm) and thickness of 250 $\mu$m. Therefore, there is a significant refractive index mismatch between the cover slip and immersion oil (RI of 1.516) of the microscope objective. The thickness of the sample solution inside the chamber is around 25 $\mu$m which can be estimated by focusing the objective lens on fiducial marks put on the cover slip and glass slide and reading off the z-travel of the objective from the vernier scale. 

\subsection{\label{fielddesign}Design of optical field}
\begin{figure}[ht]
 \centering{\includegraphics[scale=0.6]{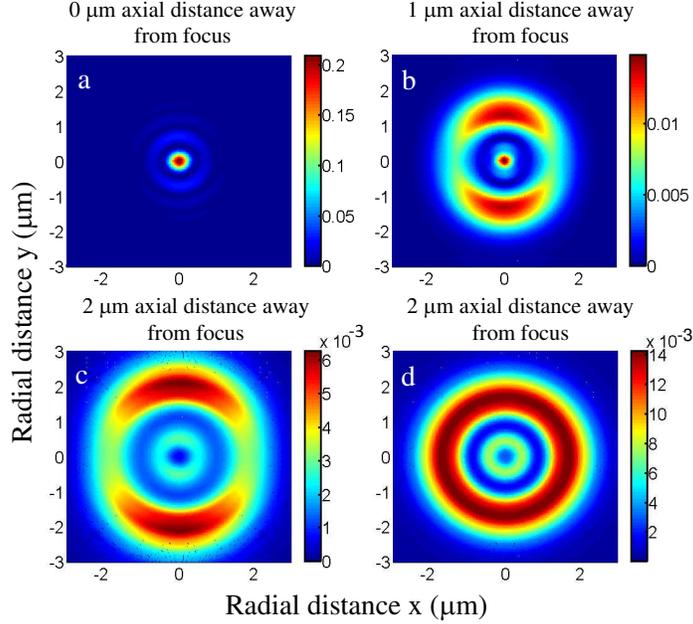}}
    \caption[radprof]{Radial profile of electric field at $x-y$ planes (a) 0 $\mu$m, (b) 1 $\mu$m, (c) 2 $\mu$m away from the beam focus, and (d) 2 $\mu$m away from the beam focus for input circularly polarized light. The sample thickness in the sample chamber is 30 $\mu$m, and the beam focus is at 21 $\mu$m inside the sample. In (a), a ring structure around the center is present, though the intensity maxima is still in the center. However, as one moves away from the focus as in (b) and (c), it is clear that the intensity maxima shifts to the ring around the center. Also, the ring diameter increases as one goes away from the focus. Radial trapping is possible anywhere in the ring, with the axial trapping stabilized by standing waves formed in the cavity defined by the cover slip and top slide. Stable radial and axial trapping can happen when an axial intensity maxima in the standing wave coincides with the radial intensity maxima as defined in the ring. Note also that the intensity inside the ring is not uniform, with two diametrically opposite regions of local maxima being formed. However, it is observed in (d) that for input circular polarization of the trapping beam, the two local maximas disappear with a continuous ring of higher intensity that is concentric with the outer lower intensity ring being formed.}
\label{radprof}
\end{figure}

In our system, the trapping laser passes through a stratified media consisting of: 1) immersion oil (RI = 1.516), 2) cover slip (RI = 1.572), 3) sample consisting of pea-pods in water solution (RI=1.33), and 4) microscope slide (RI=1.516). Thus, it experiences different RIs in each media, with the RI changing abruptly at each interface. This mismatch of different RIs', as well as the distance of travel of the beam in each media results in optical spin-orbit interaction (SOI) effects \cite{berr05, bli08, blio08, goro08, her12} which are enhanced in our case since we use both a thicker cover slip, and one which is not RI matched with the immersion oil of the objective. A comprehensive simulation of the electric field, in the axial as well as radial direction in the vicinity of the beam focus in the sample chamber, using the well-known angular spectrum method (otherwise known as vectorial Debye diffraction theory or Debye integral) \cite{Born},  led to the final expression of the electric field inside the sample chamber for both forward and backward propagating waves \cite{hal12}. This could be written in cylindrical polar coordinates ($\rho, \psi, z$) as  
\begin{equation}
\label{fieldeq}
 \vec{E_t}(\rho,\psi,z)=\left[
\begin{array}{c}
\ {E_x}\\
\ {E_y}\\
\ {E_z}\\
\end{array}
\right ]
= C
\left[
\begin{array}{c}
\ {I_0+I_2cos(2\psi)\ }\\
\ {I_2\sin(2\psi)\ }\\
\ {i2I_1 cos(\psi)\ }\\
\end{array}
\right ]
\end{equation}
where,
\begin{widetext}
\begin{eqnarray}\label{transmisseqn}
I^t_0=\int_0\limits^{min(\theta_{max},\theta_c)}E_{inc}(\theta)
\sqrt{\cos\theta}(T^{(1,j)}_s+T^{(1,j)}_p\cos\theta_j)J_0(k_1\rho\sin\theta)e^{ik_jz\cos\theta_j}sin(\theta)\>d\theta \nonumber
\end{eqnarray}
\begin{eqnarray}
I^t_1=\int_0\limits^{min(\theta_{max},\theta_c)}E_{inc}(\theta)
\sqrt{\cos\theta}T^{(1,j)}_p\sin\theta_j
J_1(k_1\rho\sin\theta)e^{ik_jz\cos\theta_j}\sin\theta\>d\theta \nonumber
\end{eqnarray}
\begin{eqnarray}
I^t_2=\int_0\limits^{min(\theta_{max},\theta_c)}E_{inc}
(\theta)\sqrt{\cos\theta}(T^{(1,j)}_s-T^{(1,j)}_p\cos\theta_j) J_2(k_1\rho\sin\theta)e^{ik_jz\cos\theta_j}\sin\theta\>d\theta
\end{eqnarray}
\end{widetext}
where we use the Fresnel coefficients $T_s$ and $T_p$ for multiple interfaces (generally complex in such cases), $J_{0,1,2}$ are standard Bessel functions, $C$ is a phase factor, while $\psi$ could be understood as the angle of polarization of the input linearly polarized Gaussian beam. It can be shown that the spatial distributions of the complex valued $I_0(\rho), I_2(\rho)$ and $I_1(\rho)$ coefficients in our system leads to enhanced SOI resulting in a polarization-dependent intensity distributions at the trapping plane for incident linearly polarized light \cite{roy12}. For the thick cover slips we use, $I_0(\rho)$ does not maximize at the beam axis, i.e. $x=0$, but slightly farther away (around $x=2~ \mu$m). Also, the value of $I_2(\rho)$ is quite large (about double of that used in standard cover slips used in tweezers), so that the resultant radial intensity distribution near the focus for sample thickness 30 $\mu$m and focus at 21 $\mu$m inside the sample ($z = 21$) becomes that shown in Fig.~\ref{radprof}.  Now, the total intensity $I(\rho)$ for incident linearly polarized light can be shown to be \begin{equation}
I(\rho) = \ \left|I_0\right|{^2} + \left|I_2\right|{^2}  \pm \ 2 {\bf Re}(I_0I_2^{\star})\cos(2\psi)
\label{intensitylinpolar}
\end{equation}
where the positive and negative signs are for horizontal ($x$-polarization) and vertical ($y$-polarization) incident polarization states respectively. Fig.~\ref{radprof} shows the radial intensity distribution plotted using Eq.~\ref{intensitylinpolar} at the beam focus, and at axial distances of 1 and 2 $\mu$m away from the focus. It is clear from the figure that even at the focus, a portion of the intensity is distributed in a ring outside the center, that gradually become stronger as one moves away from the focus, so that in Fig.~\ref{radprof}b and c, the intensity maxima  is no longer in the center but in the ring.  Also, the dependence of $I(\rho)$ on $\cos(\psi)$ as is apparent in Eq.~\ref{intensitylinpolar}, leads to the formation of diametrically opposite lobes inside the ring as is shown in Fig.~\ref{radprof}.  It is also clear from Fig.~\ref{radprof} that the diameter of the ring increases (from  around 1 to 2 $\mu$m) as one goes away from the focus. We have obtained stable trapping up to diameters of around 5 $\mu$m, after which the trap becomes too weak.  The axial trapping is stabilized by the standing wave geometry of our system (cover slip and slide acting as two plane mirrors) - interference fringes are produced in the axial direction creating intensity nodes and anti-nodes where particles can be trapped \cite{Zem01} (in the anti-nodes or maxima). Note that we had used this novel potential earlier for trapping polystyrene beads of diameter 1.1 $\mu$m \cite{hal12}, where the beads self assembled in ring patterns near the trap focus with the experimental system described above.  Another interesting exercise is to consider the effect of input circular polarization in the radial intensity profile near the trap focus. We demonstrate this in Fig.~\ref{radprof}d where we show the field intensity structure at an axial distance of 2um from the focus for input circular polarization. The local maxima is now smeared out all along the background intensity ring, so that transport of particles by changing the input polarization is not possible. However, due to conservation of angular momentum, there would an orbital angular momentum associated with the field that the particle could experience. We are presently carrying out investigations in this direction. 

\begin{figure}[ht]
 \centering{\includegraphics[scale=0.6]{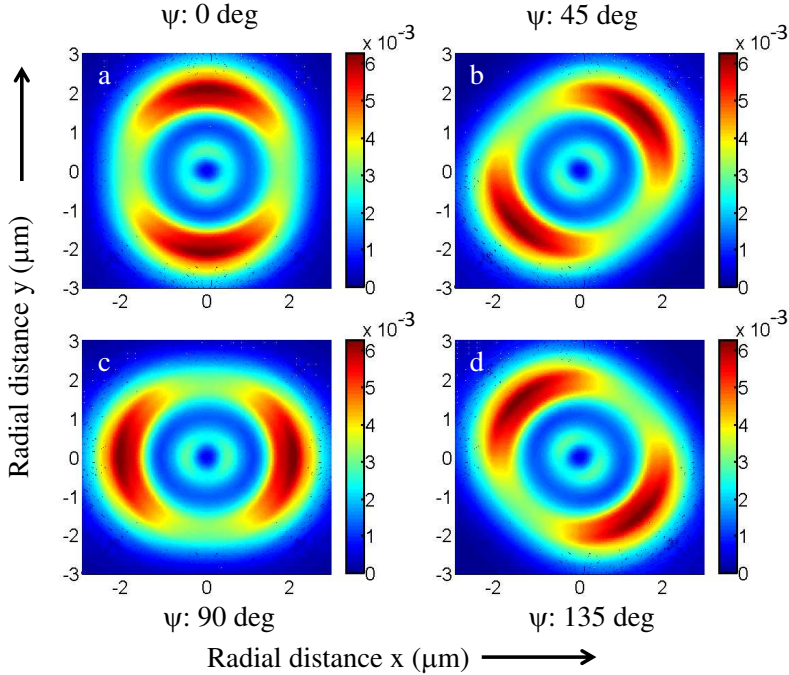}}
    \caption[]{Radial intensity profile of electric field at an $x-y$ plane 1 $\mu$m away from the beam focus. The sample thickness and beam focus position are same as in Fig.~\ref{radprof}. The plots are for different polarization angles of the input Gaussian beam with the polarization angle $\psi$ equal to (a) 0 degrees, (b) 45 degrees, (c) 90 degrees, and (d) 135 degrees. It can be seen that the regions of local maxima inside the high intensity ring  can be rotated along the ring by changing the value of $\psi$. Thus, a particle trapped in a local maxima can be translated along the circumference of the ring.}
\label{Figure-2}
\end{figure}

Single pea-pods can be preferentially trapped in the local intensity maxima inside the ring, and can then be transported along the periphery of the ring by changing the polarization angle ($\psi$) of the linearly polarized trapping beam at the input of the trap. The variation of the location of the local maxima inside the ring can be seen in Fig.~\ref{Figure-2}, where sub-plots (a), (b), (c), and (d) are drawn for input polarization angles ($\psi$) of 0, 45, 90, and 135 degrees respectively. This is how we achieve controlled motion of pea-pods in the optical trap.

\subsection{\label{synth}Synthesis of SOM pea-pods and experiments to test its catalytic actions}
The synthesis used here is an optimization over the synthesis reported by us earlier \cite{sou07}. Ammonium phosphomolybdate, ${\rm (NH_4)_3[PMo_{12}O_{40}]}$ (MW - 1876.35 g/mol), was used without further purification. The dispersion of ammonium phosphomolybdate in water was prepared in the following manner: To 10 mg of ammonium phosphomolybdate 5 mL of distilled water was added. The suspension was sonicated for 30 min, and the undissolved remnants of yellow phosphomolybdate Keggin were removed by filtration using standard Whatman filter papers. The filtrate was allowed to stand undisturbed for 3 more days after which pea-pod shaped structures were seen by AFM and later used for optical trapping experiments. To this dispersion of SOM peapods, four different concentrations of ${\rm(NH_4)_3[PMo_{12}O_{40}]}$ were loaded. The catalyst loaded dispersions were characterized by AFM and TEM and those dispersions were used for catalyzing oxidation of benzaldehyde to benzoic acid as described below. The AFM used was a NT-MDT, NTEGRA system with z-resolution ~1 nm. The size distribution of pea-pods thus obtained varies from 1 - 3 um (long axis).

In a 20 ml microwave reaction tube, benzaldehyde (0.5 mmol) dissolved in acetonitrile (5 ml), were mixed with 30\% hydrogen peroxide (0.25 ml) to which 4 different dispersions of SOM peapods loaded with ${\rm (NH_4)_3[PMo_{12}O_{40}]}$ were added. The contents of the flask were sealed and were irradiated for 10 minutes at 75$^0$C under constant stirring. After cooling, acetonitrile was evaporated on vaccum-evaporator. The aqueous layer was extracted with ethyl acetate ($3\times 10$ ml). The combined organic layer was washed with brine, filtered and concentrated in vacuum. The crude reaction mixture gave benzoic acid by flash column chromatography on silica gel. The as-obtained benzoic acid was further purified by re-crystallization from its aqueous solution, its weight measured and the yield recorded. 

\section{\label{resdisc}Results and discussions}
\subsection{\label{peatrap}Trapping pea-pods and finding diffusion coefficient} 
\begin{figure}[ht]
 \centering{\includegraphics[scale=0.6]{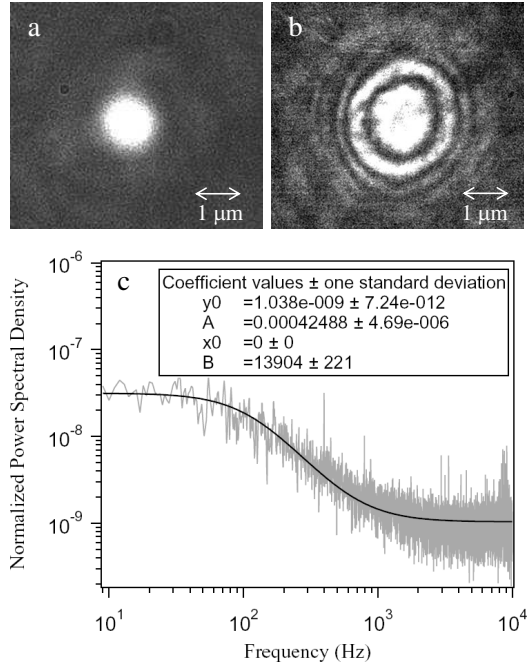}}
    \caption[powerpea]{(a) Trapping laser intensity at the focus of the optical trap imaged using the back-scattered light that is incident on the CCD camera attached to the microscope side port. The intensity distribution shows a Gaussian structure, which matches the simulation results shown in Fig.~\ref{radprof}a. The spot size is around 1 $\mu$m. (b) Intensity distribution imaged at a plane about 2 $\mu$m away from the focus. The ring structure as predicted by the simulation results shown in Fig.~\ref{radprof}b is visible. The spot size is around 2 $\mu$m. (c) Typical power spectra obtained using our detection system for a trapped pea-pod of dimension around $2 \times 0.5 \mu$m. The power spectrum fits well to a Lorentzian yielding a corner frequency (given by square root of fit parameter B) of around 118 Hz, yielding a trap stiffness of about 12.5 pN/$\mu$m. The value of the fit parameter $A$ which gives the diffusion constant $D$ comes out to be 0.00042.}
\label{powerpea}
\end{figure}

Figs.~\ref{powerpea}a and b show how the trapping laser spot size varies both in size as well as structure as we change the z-focus of the microscope. These images have been taken with the CCD camera attached to the side port of our microscope. It should be noted, however, that the imaged field intensity will be a superposition of the intensity at all planes in the sample, and reflected intensities from the different chamber surfaces, and may not be exactly indicative of the intensity distribution at the particular plane of interest. It is apparent that the imaged spot is almost a clean Gaussian at the trap focus (Fig.~\ref{powerpea}a) of size close to 1 $\mu$m, while a larger ring structure of size around 2 $\mu$m is visible as we move the z-focus of the microscope by around 2 $\mu$m (Fig.~\ref{powerpea}b). Initially single pea-pods are trapped at the focus (simple Gaussian structure). Once trapped, we determine the diffusion coefficient of the pea-pods in order to ascertain that the optical trap is sufficiently harmonic near the center,  and that we could use it to probe the dynamics of the pea-pods in solution accurately. It is well known that an optically trapped particle would execute only uncorrelated Brownian motion at the trap center, the power spectrum of which would be a Lorentzian \cite{berg04}. A Lorentzian fit to the power spectrum of the Brownian motion yields the diffusion coefficient, which could then be compared to the diffusion coefficient measured independently earlier \cite{sou07}. Now, an optically trapped particle executing Brownian motion obeys a simplified Langevin equation, so that the power spectrum $P_k$ can be written as \cite{berg04}
\begin{equation}
\label{powrspc}
P_k = \frac{D/(2\pi^2)}{f_c^2 + f_k^2}
\end{equation}
where $f_c$ is the corner frequency defined as
\begin{equation}
\label{f_c}
f_c\equiv\kappa/(2\pi\gamma_0)
\end{equation}
and
\begin{equation}
\label{diffeqn}
D=k_BT/\gamma_0
\end{equation}
is the diffusion coefficient, with $k_B$ being the Boltzmann constant, and $T$ the temperature (298K). $\gamma_0$ is the coefficient of friction given by $\gamma_0 = 6\pi a \beta$, where $a$ is the radius of the particle, and $\beta$ is the coefficient of dynamic viscosity of the fluid medium in question (water in our case, and we have assumed $\beta$ = 898 mPa), while $\kappa$ is the spring constant (stiffness) of the harmonic trap. A typical power spectrum of a trapped pea-pod is shown in Fig.~\ref{powerpea}. 

The corner frequency comes out to be about $118 \pm 2$ Hz, from which we obtain 
a value of the trap stiffness $\kappa$ of 12.5 pN/$\mu$m using Eq.~\ref{powrspc} for a trapped pea-pod of dimension around $2 \times 0.5 \mu$m. Also, as shown in Fig.~\ref{powerpea}, another fit parameter that emerges out of the fitting procedure is $A$. This is related to the diffusion constant $D$ of the Einstein theory by the relation \cite{svo04}
\begin{equation}
\label{diffpara}
D = \dfrac{2\pi^2 A}{\rho^2},
\end{equation} where $\rho$ is the normalized displacement sensitivity of our detector ($\sim 0.16(1) ~\mu m^{-1}$) for light at 532 nm scattered off the pea-pods. Using a fit value of 0.00042, we obtain $D$ as $3.2(5)\times 10^{-13} {\rm m^2/s}$, which is in very close agreement with the value of $3.3\times 10^{-13} {\rm m^2/s}$ reported earlier \cite{sou07}. This gives us confidence that the optical trap is well-characterized and could be used reliably to measure dynamics of the pea-pod in solution.

\subsection{\label{peamoved} Controlled motion of a single pea-pod}  
\begin{figure*}[ht]
 \centering{\includegraphics[scale=0.65]{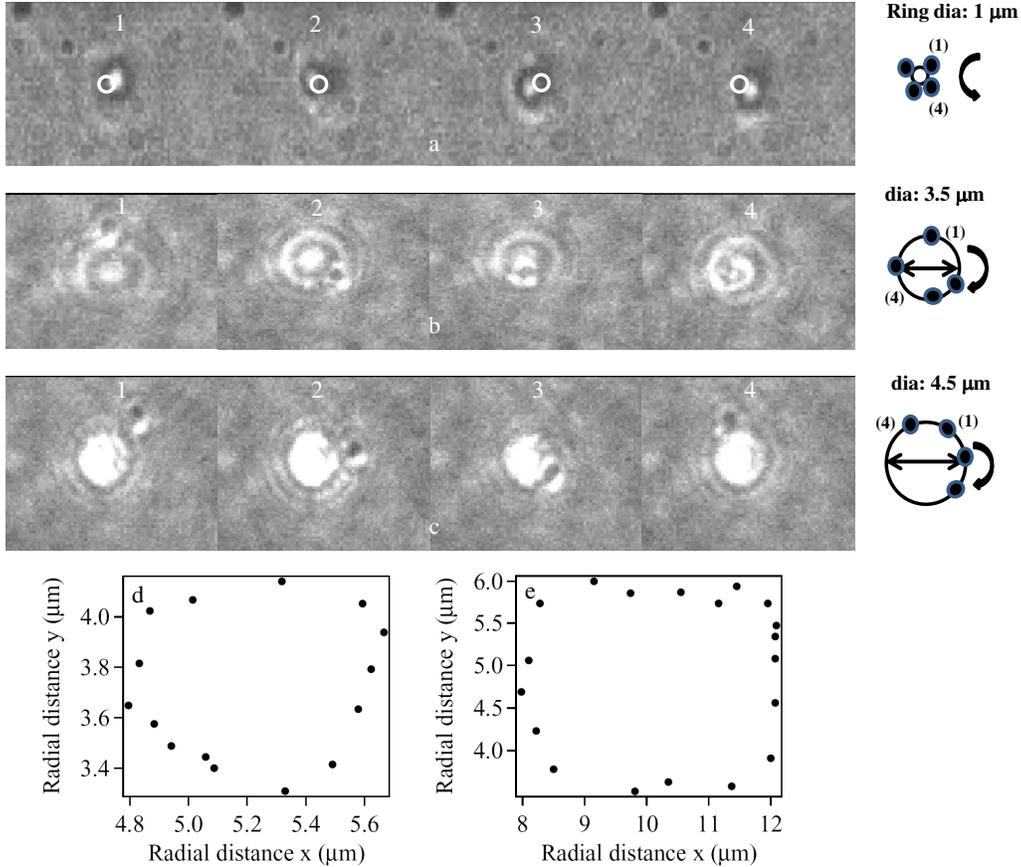}}
    \caption[peamove]{Controlled motion of a single optically trapped pea-pod along the periphery of the intensity maxima ring. a, b and c show time series of four positions of single pea-pods along the periphery of rings having diameter 1, 3.5, and 4.5 $\mu$m respectively. For b and c, the trapping light profile is visible indicating the pea-pod orbit. In a, a filter at the camera input blocks the light and we indicate the orbit by a white circle. Each pea-pod position has been obtained by rotating the linear retarder at the input of the optical trap so as to change the polarization angle of the trapping beam. The pea-pod can be moved along the entire periphery by 360 degrees. Subplots d and e show the quantified values of the particle trajectories as obtained from a commercial particle tracking software. A ring diameter of around 0.8 $\mu$m is obtained for d, while that of around 4 $\mu$m is obtained for e.}
\label{peamove}
\end{figure*}

We then proceed to move a single trapped pea-pod in a controlled manner. As shown in Fig.~\ref{peamove}, the pea-pod is trapped in the ring just outside the centre of the beam by controlling the microscope z-focus. It is trapped in a local intensity maxima in the ring, and as had been demonstrated in Fig.~\ref{Figure-2}, we change the input polarization so as to cause the local maxima to move along the periphery of the intensity maxima ring. This causes the pea-pod to be transported controllably along the ring. The roughly spheroidal shape of the pea-pods leads to the existence of a unique symmetry axis which helps them to align in the direction of the gradient of the electric field intensity that, in this case, would be tangential to the intensity ring. It is interesting to note that particles which are much smaller in dimension than the extent of the local intensity maxima in the ring would not possibly orient along the tangent since they are too small to see the overall direction of the intensity gradient. The orientation of the peapods are apparent in Frames 2 and 3 in Fig.~\ref{peamove}b, and Frames 1, 2, and 3 in Fig.~\ref{peamove}c where the orientation of the peapod changes as it traverses different regions of the intensity ring (following the direction of the intensity gradient that is always tangential to the ring). Also, the pea-pods being soft oxometalates have low effective hydrodynamic mass \cite{berg04}, which reduces the  frictional drag force due to the medium so that they can be easily dragged by the field gradient produced when the input polarization is modified. We demonstrate three different orbit diameters corresponding to 1, 3.5 and 4.5 $\mu$m in Figs.~\ref{peamove} a, b, and c respectively. To measure the diameters of the rings traversed by the pea-pods, we use a particle tracking software, where multiple images of the particle in motion are fed into a commerical software 'Cell Professional'. The software finds out the location of the particle by checking the contrast with respect to the background and making a non-linear fit to the cross section. Spurious points are filtered by specifying proper threshold values and the minimum area of the detected region in the image. Subsequent positions of the target particle are tracked in a similar manner. The positions of the centroids of the particle's image are then plotted as shown in Fig.~\ref{peamove}d and e, where diameters of around 0.8 and 4 $\mu$m are obtained for the particle motion, respectively. The corresponding real time videos of particle motion in all the above cases are available in the Supplementary Material \cite{supplem}.

We can trap SOM pea-pods in rings of diameter between 1 - 5 $\mu$m, corresponding to periphery lengths of 3 - 16 $\mu$m. The stiffness of the optical trap reduces as the ring diameter is increased (since the field intensity is reduced), so that beyond 5 $\mu$m, the trap becomes quite weak and we often lose the particles due to the torque generated when the particles are moved in the circular orbit. Pea-pod sizes are also important, for such controlled motion and pea-pods with lengths less than $2 \times 1~ \mu$m can be moved easily. 

\subsection{\label{peacatal}SOM peapods in catalysis}
\begin{figure*}[!h!t]
 \centering{\includegraphics[scale=0.65]{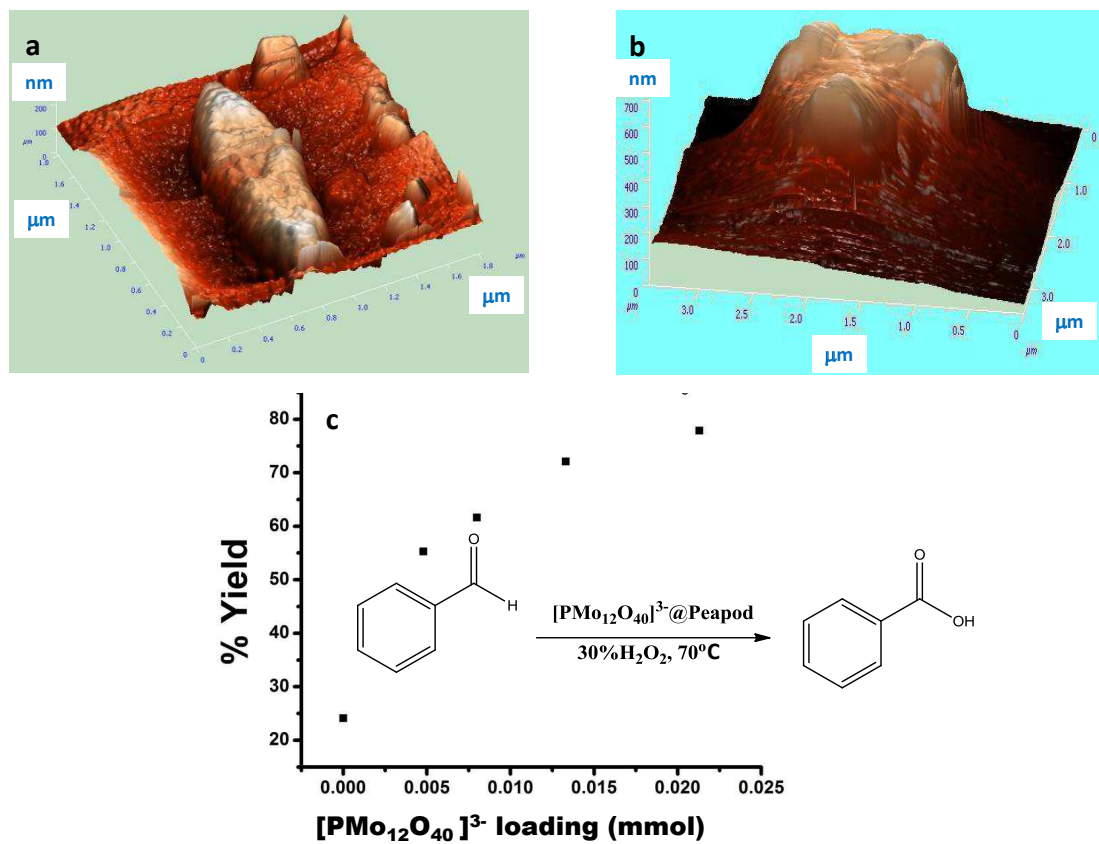}}
    \caption[peacatal]{SOM Peapods as catalyst carrier. (a) shows the AFM image of a single peapod, (b) shows clear change in topology and dimensions of ${\rm [PMo_{12}O_{40}]^{3-}}$ loaded catalyst and (c) shows the catalytic activity of the catalyst as function of catalyst loading on SOM peapod carrier from a spectroscopic analysis. The catalyzed reaction is also shown schematically.}
\label{peacatalafm}
\end{figure*}

We now proceed to study the role of SOM peapods in catalysis. To do so we investigate the catalyst carrier capacity of SOM peapods. Since the SOM peapods have skins of $[MoO_3]$ -  a versatile catalyst carrier, it is reasonable to believe that the peapods would have similar capacity. Indeed we observe that it is possible to load ${\rm [PMo_{12}O_{40}]^{3-}}$ on the surface of the peapods as observed from the clearly corrugated and thickened surface of catalyst loaded peapods that are shown in an AFM image (Fig.~\ref{peacatalafm}(b)). It is clear from the AFM images (Fig.~\ref{peacatalafm}(a) and (b)) that the height of a standard peapod after loading with ${\rm [PMo_{12}O_{40}]^{3-}}$ increases by almost a factor of three (200 to 600 nm), while the length increases by around 45\% (1.8 $\mu$m to 2.5 $\mu$m). The other particles visible in Fig.~\ref{peacatalafm}(b) are ${\rm[P_2MoO_{11}]}$ spherulites and have been described in detail elsewhere \cite{sou11}. Note that the AFM image is representative and a similar trend is obtained  after imaging several peapods before and after treatment with the catalyst carrier. Moreover, to check the catalytic activity of this catalyst loaded peapod dispersion, we now use this dispersion for catalyzing oxidation of benzaldehyde to benzoic acid. Clearly we observe a significant effect of catalyst loading on the yield of benzoic acid, which was tested gravimetrically and spectroscopically by $^1$H NMR spectroscopy. The results have been shown in Fig.~\ref{peacatalafm}(c). Clearly the loaded SOM peapod dispersions catalyze the oxidation reaction and the yield of the product increases with an increase in loading. The loading is also irreversible. The yield increases significantly up to 75\% upon catalyst loading as compared to the situation when no catalyst in present. The activity of SOM peapods as catalyst carriers and hence in catalysis is thus conclusively demonstrated. Also, since the $\rm{[PMo_{12}]}$ Keggins used are not soluble in water, one can safely conclude that catalysis happens on the pea-pods only and not in solution. 

\section{\label{concl}Conclusions}
To summarize, we have demonstrated controlled motion of single SOM pea-pod along the periphery of an optical trap that has a unique ring-like structure due to effects of SOI. Pea-pod displacements of more than 16 $\mu$m have been obtained without changing the optical trap stiffness during the motion by simply controlling the polarization angle of the trapping beam incident upon the microscope objective. Higher displacements along more complex paths are also achievable with the use of  a radial array of overlapping intensity rings (each in itself an optical trap) by coupling multiple beams into the trapping chamber. A pea-pod trapped in one of the rings may be brought at the intersection region of the adjacent ring by varying the input polarization angle, after which the first ring (beam) could be switched off so that the pea-pod is trapped in the adjacent ring. The process can be repeated for multiple rings so that the pea-pod can be moved in circular or semi-circular paths while also being translated radially in the x or y direction. Displacements of several hundred microns are possible to be envisaged in such configurations. The rotation of these SOM-pea-pods paves the path in providing a means of deliberated information transfer from a single molecular level to micro- and macroscopic level using simple and controllable physico-chemical methods with potential implications in controlled micro-reactors, drug-delivery and related nano-systems. It may be noted in passing that the SOM-pea-pod used here is a potentially high surface area mesoscopic catalyst carrier owing to the presence of ${\rm MoO_3}$ on its entire surface and has also been shown here by us. We demonstrate the such capacity by catalyzing a model reaction of oxidation of benzaldehyde to benzoic acid using $[PMo_{12}O_{40}]^{3-}@Peapod$ as catalyst. The results clearly show the ability of SOM peapod as catalyst carrier. Hence such a SOM based MOM reported here could be of use for trafficking catalysts from a region loaded with `selective' reagents to that loaded with `reactive' reagents and thereby enhance catalytic activity immensely. Needless to say, the library of mesoscopic SOMs is vast, emerging and expanding. MOMs with SOMs could pave a new path for designing of newer interactive systems tailored to meeting new as well as existing needs of our times.       
\section{Acknowledgements}
This work was supported by the Indian Institute of Science Education and Research, Kolkata, an autonomous research and teaching institute funded by the Ministry of Human Resource Development, Govt. of India. This paper is dedicated to Prof. Jean-Marie Lehn. 

\nocite{*}

\providecommand{\noopsort}[1]{}\providecommand{\singleletter}[1]{#1}

\end{document}